\documentclass[journal, letterpaper.twocolum,a4paper]{IEEEtran}
\IEEEoverridecommandlockouts
\usepackage{cite}
\usepackage{amsmath,amssymb,amsfonts}
\usepackage{dsfont}
\usepackage{graphicx}
\usepackage[export]{adjustbox}
\usepackage{algpseudocode}
\usepackage{textcomp}
\usepackage{bbm}
\usepackage{xcolor}
\usepackage{booktabs}
\usepackage{subcaption}
\usepackage{dblfloatfix}
\usepackage{soul} 
\usepackage{comment}
\newcommand{\comm}[1]{}
\usepackage[ruled,vlined,linesnumbered]{algorithm2e}
\usepackage{multirow}
\def\BibTeX{{\rm B\kern-.05em{\sc i\kern-.025em b}\kern-.08em
    T\kern-.1667em\lower.7ex\hbox{E}\kern-.125emX}}
\usepackage{amsmath}

\newtheorem{remark}{Remark}
\newcommand{\red}{\textcolor{red}}

\newcommand{\blue}{\textcolor{blue}}
\newcommand{\adapt}{AdaptSky}
\newcommand{\pro}{\ensuremath {\tt{ProSky}}}

\newcommand{\mmwave}{mmWave}
\newcommand{\uav}{UAV}
\newcommand{\uavs}{UAVs}

\newcommand{\nue}{\ensuremath {\tt{N_{UE}}}}
\newcommand{\ue}{\ensuremath {\tt{UE}}}
\newcommand{\mimo}{\ensuremath {\tt{MIMO}}}

\newcommand{\noma}{NOMA}

 \usepackage{geometry}
 \usepackage{enumitem} 
\IEEEoverridecommandlockouts\IEEEpubid{\makebox[\columnwidth]{ 978-1-6654-5975- 4/22~\copyright~2022 IEEE \hfill} \hspace{\columnsep}\makebox[\columnwidth]{ }}
\begin{document}

\title{
   NEAT Beating DRL in Managing NOMA-mmWave-\uav~Resources
}
\title{
 How Efficiently   Can NEAT Manage NOMA-MmWave-UAV  Resources?
}
\title{
  Can NEAT Manage NOMA-MmWave-UAV  Resources Efficiently?
}
\title{
 How Efficient   Can NEAT Be in Managing NOMA-MmWave-UAV  Resources?
}
 \title{
 ProSky:  NEAT Meets  NOMA-mmWave \\ 
 in the Sky of 6G  
 }
 
\author{Ahmed~Benfaid, 
        Nadia~Adem, and Abdurrahman Elmaghbub 

\author{\IEEEauthorblockN{ Ahmed Benfaid$^*$, Nadia Adem$^*$, and Abdurrahman Elmaghbub$^{**}$}\\
\IEEEauthorblockA{$^*$University of Tripoli, Tripoli, Libya, E-mail: \{a.benfaid,n.adem\}@uot.edu.ly}\\
\IEEEauthorblockA{$^{**}$Oregon State University, OR, USA, Email: elmaghba@oregonstate.edu
\vspace{-3.5em}
} 
}
\comm{
\thanks{Ahmed Benfaid and Nadia Adem are with the Department of Electrical and Electronic Engineering, University of Tripoli, Libya (e-mail: fbenfaid@gmail.com and n.adem@uot.edu.ly). Abdurrahman Elmaghbub is with the School of Electrical Engineering and Computer Science, Oregon State University, USA (e-mail: elmaghba@oregonstate.edu)}

\vspace{-2.5em}
}
}
\maketitle
\begin{abstract}

Rendering to their abilities to  provide ubiquitous connectivity, flexibly and cost effectively,  
 unmanned aerial vehicles   (\uavs)   have  been  getting more and more  research attention. 
 %
 To take the \uavs\textquotesingle~performance to {the} next level, however, they need to be merged with some other technologies like
  non-orthogonal multiple access (\noma) and  millimeter wave  (\mmwave), which both promise  
 high spectral efficiency (SE). 
    As managing  \uavs~efficiently may not be possible using   {model-based} 
  techniques, another key innovative technology that \uavs~will inevitably need to leverage is artificial intelligence (AI).   %
   Designing an AI-based technique that adaptively allocates radio resources and place{s} \uavs~in 3D space to meet  certain communication objectives, however, is  a tough row to hoe.  
   In this paper, we propose a neuroevolution of augmenting topologies NEAT  framework, 
   referred to as \pro, to  manage   NOMA-mmWave-\uav~networks.
   \pro~exhibits a  remarkable   performance improvement over a model-based
  method. Moreover, \pro~learns   $5.3$ times faster than and outperforms, in both SE and energy efficiency EE while being reasonably fair, a deep reinforcement learning DRL based scheme. The \pro~source code is accessible to use here: https://github.com/Fouzibenfaid/ProSky
  
   \comm{Sufficient numerical results show that our proposed strategy has a remarkable advantage over existing systems in terms of energy efficiency.}

\end{abstract}

\begin{IEEEkeywords}
   Deep reinforcement learning (DRL), 
   millimeter wave (mmWave),   neuroevolution of augmenting topologies (NEAT), non-orthogonal multiple access (NOMA), 
   unmanned aerial vehicle (\uav).  
\end{IEEEkeywords}

\vspace{-1em}
\section{Introduction} 
\label{sec: Introd}
Despite the unprecedented advancements in telecommunication technologies in recent years, around half of the world\textquotesingle s population, mostly living in rural and developing areas, still has limited or no access 
to cellular communication services\cite{yaacoub2020key}. 
Providing connectivity to those underprivileged areas could unequivocally enhance the quality of their 
{lives}. 
One of the envisioned, must-be-met, requirements of 
 6G 
is 
enabling ubiquitous geographical coverage 
 anywhere, anytime.
Due to the lack of essential cellular infrastructures in rural areas and the high cost of establishing them, along with the incapability of terrestrial base stations (BSs) to cover hotspot areas during special events or disaster scenarios, unmanned aerial vehicles (\uavs), thanks to their flexible 3D mobility and ease of deployment, 
are envisioned to be a major part of the 6G wireless networks, acting as flying BSs~\cite{adem2021crucial}. 
 Allowing sharing same spectrum resources among multiple wireless nodes simultaneously, non-orthogonal multiple access (NOMA)   is emerging as another 6G enabler 
 offering massive device connectivity without exhausting spectrum resources~\cite{adem2021crucial}.
  {The emergence of NOMA-aided \uav~networks necessitates a careful investigation of  optimal UAV 3D  deployment, and power allocation (PA) management~\cite{adem2021crucial}}. 
  Due to the high complexity of such an optimization problem, authors in~\cite{Chen2019}, for example, approached the \uav~placement and PA problems disjointly, whereas the number of served users was limited to two in~\cite{Monemi2020,Sharma2017},
 and the \uav~mobility was restricted to a 2D plane in~\cite{Monemi2020,Chen2019,Sharma2017}. 
 %
  {There have been some attempts to address the UAV 3D placement problem, but these have primarily been accomplished by disjoining the UAV placement and PA problems, as done in \cite{el2019learn}.}
 %
%
 
 Furthermore, due to the high probability 
 {of} line-of-sight (LoS) links UAVs offer, 
  the vast millimeter wave (mmWave)  and terahertz  spectrum can be utilized to satisfy the major 6G data rate enhancement requirement~\cite{adem2021crucial}. 
Nevertheless, managing NOMA-mmWave-\uav~network resources to satisfy dynamic, heterogeneous, and massive needs adaptively yet fairly and efficiently is a complicated challenge, which conventional and even machine learning methods fail to handle. 
The good news, however, is that deep learning and, more generally, artificial intelligence (AI) technologies  have the strong potential to handle multi-state network statuses and demands.  After proving their effectiveness in solving problems with a large degree of freedom  in various fields,  AI techniques are proposed to be a key enabler for self-organizing, self-optimized networks in the 6G era~\cite{adem2021crucial}.
%
 
 Inspired by the remarkable successes of incorporating deep reinforcement learning (DRL) into different fields, a DRL model has been used to solve the placement problem of  \uavs~that fly at a fixed height~\cite{liu2018energy}. More interestingly, our previous work, AdaptSky~\cite{adapt},  unlike any other work, jointly solved the non-convex optimization problem 
  of the 3D deployment and the PA of a NOMA-equipped \uav~BS in the mmWave spectrum. 
Another AI tool that {has} recently 
shown outstanding performance in a  variety of applications, including robotics and gaming~\cite{stanley2019designing}, is  neuroevolution of augmenting topologies (NEAT)~\cite{stanley2002evolving}. 
Furthermore,  although fairly limited,  NEAT  has been recently used in the area of communications, for example, in~\cite{kang2021neuroevolution} to improve  beam management in vehicle-to-vehicle communications.   
     To the best of our knowledge, no work has studied the use of NEAT to manage  UAV-based communications.
 In this work, we propose \pro, a novel NEAT-based framework that jointly optimizes 3D deployment and PA  for NOMA-aided \uav~BSs operating in the mmWave spectrum. 
 The  main contributions of this work, and advancements over~\cite{adapt},  are summarized as follows:  
\begin{enumerate}[label=(\roman*)]
\item   \pro~incorporates and demonstrates the effectiveness of  NEAT, for the first time, to manage NOMA-mmWave-\uav~networks.  We set the NEAT environment that leads to the optimal \uav~placement and NOMA PA such that, without sacrificing fairness, the total network data rate is maximized. 
\item Although DRL based algorithms show  tremendous improvements over a state-of-art  in managing 3D networks while exhibiting high generalization capabilities~\cite{adapt}, they have two drawbacks:    i) specifying their neural network (NN) structure beforehand 
and tuning it using  trials and errors does not only lead to degradation in efficiency but also some times in  performance,   ii) they train, relatively, slow and run into  local minima issues.   As NEAT {optimally}  determines the NN structure, 
 \pro, however, demonstrates, while being reasonably fair,  $5.8\%$ improvements in  SE and  $21.9\%$  in EE  over a DRL based benchmark.  More interestingly, as NEAT  simultaneously trains multiple NNs {to}  evolve based on a genetic algorithm (GA),  \pro~exhibits more than $400\%$ improvement in learning rate.
\end{enumerate}

\comm{
\subsection{Organization}
\label{subsec: Organiz}
The rest of this paper is organized as follows. The system model and problem formulation are presented in Section \ref{sec: Sys & prob}. AdaptSky framework is proposed in Section \ref{sec: adaptsky}. In Section \ref{sec: numr}, we provide AdaptSky simulation results. Discussions and outlook is presented in Section \ref{sec: disc}. Finally,  conclusion\st{s}  \st{are} is drawn in Section \ref{sec: conc}. 
}
\vspace{-.5em}
\section{System model}
\label{sec: Sys Model} 
\subsection{Network Model} 
\label{subsec: Network} 
\begin{figure}[t]
    \centering \includegraphics[width=0.5\textwidth]{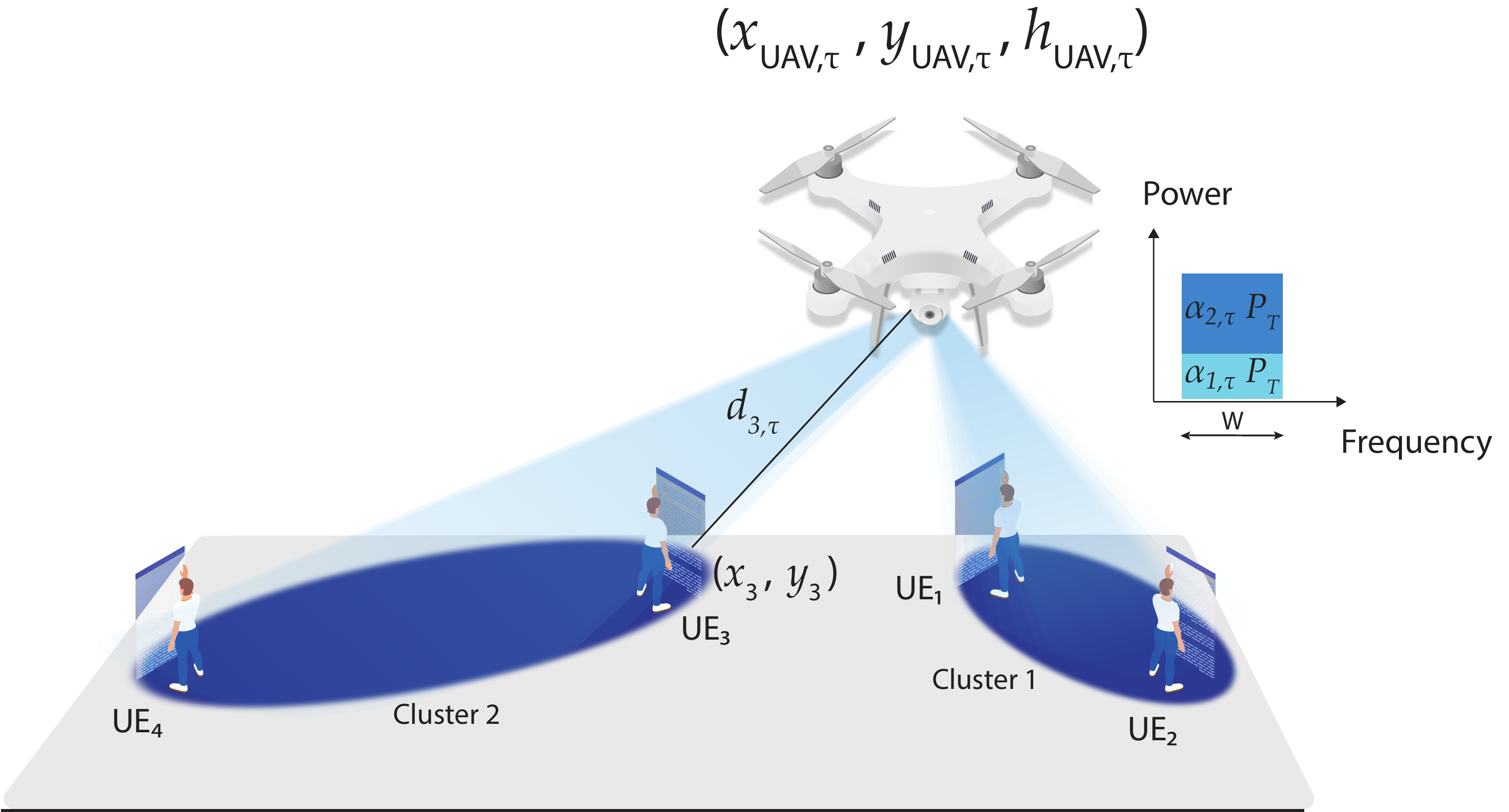}
    \caption{System model.}
    \label{fig:system_model}
\end{figure}

Embracing the 3D coverage capability of \uav s, and similar to our work in~\cite{adapt}, we consider a 3D downlink cellular network that covers an area $\mathcal{A}$ of  
 $L \times L$ units in which a \uav~serves a total of $2{N}$, for some integer number ${N}$, uniformly distributed ground users, $ \nue$. The \uav~and ground users are assumed to be  equipped with $\mathcal{N}_{\uav}$ and $\mathcal{N}_{\ue}$ antennas, respectively. Throughout the paper, user $i$ is denoted by ${\ue}_i$ where $i \in \{1,2,..,\nue\}$.   
 We assume that the users are grouped into clusters, as depicted in Fig.~\ref{fig:system_model}, in such a way that users ${\ue}_i$  and ${\ue}_{i+1}$  for $i \in \{1,3,..,\nue-1 \}$  are associated with the same cluster and 
 $\ue_i$ has a stronger channel gain than ${\ue}_{i+1}$, 
 following the distance-based pairing strategy discussed in~\cite{liu2019uav}. The assumption of having $2{N}$ users is set only for convenience and should not affect the model\textquotesingle {s} generality. In case there is {an} odd number of users, a cluster will encompass a single  user and every thing else is still valid. 
The \uav~serves each cluster over an orthogonal power resource with a total power $P_T$,  distributed between the two corresponding users based on their channel conditions.
 The received power 
at the ${\ue}_i$  
at a given time  step $\tau$ can be expressed as
\begin{eqnarray} \label{eqn: output}
 \hat{P}_{i,\tau}  &=& P_T{g}_{i,\tau}^{\mimo}(d_{i,\tau})  {\alpha}_{i,\tau} ,
\end{eqnarray}
where ${g}^{\mimo}_{i,\tau}(d_{i,\tau})$ is {the gain}  
between the \uav~and ${\ue}_i$  separated
by a 3D distance of $d_{i,\tau}$, and ${\alpha}_{i,\tau}$ is the percentage of the $P_T$ assigned to ${\ue}_i$. 
If we consider only a large scale fading 
and assume a slight difference between antenna pairs, the  gain can be approximated as ${g}^{\mimo}_{i,\tau}(d_{i,\tau}) = G {g}_{i,\tau}(d_{i,\tau})$, {where  $G$, equals $ \mathcal{N}_{\uav} \times$ $\mathcal{N}_{\ue}$, is the  gain resulting from applying multiple-input multiple-output (MIMO) antenna configurations at the  \uav~and ${\ue}_i$. 
 ${g}_{i,\tau}(d_{i,\tau})$ is the channel gain.
}
%

Based on the principle of NOMA, the superposition coding (SC) is used  at the \uav~to transmit  
signals to  users located in the same cluster. SC encodes different signals into a single signal while assigning them different power values.
 The successive interference cancellation (SIC) is used at the receiver side for signal detection. The received signal to interference plus noise ratio   at $\tau$ for ${\ue}_i$, SINR$_{i,\tau}$, is expressed as 
\begin{equation} \label{eqn: SINR}
SINR_{i,\tau}  = 
\frac{P_T G {g}_{i,\tau}(d_{i,\tau}) \alpha_{i,\tau} }{P_T  G {g}_{i,\tau}(d_{i,\tau}) \beta_{i,\tau}  + \sigma^2}
\end{equation}
%
where $\beta_{i,\tau}=\alpha_{i-1,\tau}$ if $i$ is even and zero otherwise. 
  $\sigma^2$ is the noise power.
  The first term in the denominator of \eqref{eqn: SINR} represents the interference from the user with the stronger channel gain on the other user in the same cluster. \comm{Based} Using \comm{on} the SIC technique, however, the interference {from  the user with the stronger channel gain gets}  removed. 
  %
 The data rate of ${\ue}_i$ at $\tau$ is given by
 \vspace{-0.5em}
\begin{equation} 
\label{equ:ith-user_rate}
R_{i,\tau}  = W\log_2(1+SINR_{i,\tau} ),
\end{equation}
where $W$ is the communication bandwidth. 
\label{subsec: Channel} 
     The channel gain, ${g}_{i,\tau}(d_{i,\tau})$, between the \uav~and $\ue_i$  at the  mmWave spectrum, in the presence of the LoS link~\cite{li2018uav},
is expressed as~\cite{akdeniz2014millimeter}
\begin{equation} \label{eqn: mmWave-pl}
 {g}_{i,\tau}(d_{i,{\tau}}) = C d_{i,\tau}^{-a},
\end{equation}
where $a$  and $C$ are the  path loss exponent and  intercept,  respectively. 
We  assume that the \uav~collects the channel state information  
at the beginning of each time step~\cite{el2016power}, in that
  a user can send a pilot signal prior to transmission to allow the \uav~to estimate $g_{i,\tau}$.
\vspace{-1em}
\subsection{\uav~Mobility Model}
%
The \uav~is assumed to be located at $(x_{\uav,\tau}, y_{\uav,\tau}, h_{\uav,\tau})$  at time step $\tau$, and it is able to move, in the next time step, to 
 $(x_{\uav,\tau}+d_x \delta_x , y_{\uav,\tau}+d_y \delta_y , h_{\uav,\tau}+ d_h \delta_h)$, where  $d_{x}$, $d_{y}$, and $d_{h}$ $\in\{1,-1\}$. 
 $\delta_x$,  $\delta_y$, and $\delta_h$  are the magnitude of change  in the $x$, $y$, and z \comm{axis} axes, respectively.  $h_{\uav,\tau}$ is assumed to have a minimum value of $h_0$. 
 
 \comm{
 We assume that at  {$\tau=0$}, the \uav~is located at $(0,0,h_{\uav,0})$, where $h_{\uav,{0}}$ is the initial height.} 
 %
%
%


\section{3D \uav~Placement and Power Allocation Formulation} 
  \label{sec: Problem} 
We propose to optimize the \uav~placement and PA such that the total users\textquotesingle~data rate is maximized subject to a fairness condition imposed through satisfying a minimum rate for each user, $R_{min}$. 
We define the total  users\textquotesingle~data rate at   $\tau$ as
 \vspace{-0.5em}
\begin{equation}
    R_{\tau}^{\textrm{tot}} =  \sum_{i=1}^{\nue} W\log_2(1+SINR_{i,\tau}).
\end{equation}
\vspace{-1em}


 The   optimization problem is formulated as  follows 
\begin{subequations}\label{eq:alloc}
\begin{align}
&\max_{x_{\uav,\tau},y_{\uav,\tau},h_{\uav,\tau},\alpha_{i,\tau}}   R_{\tau}^{\textrm{tot}},  \label{eqn:objective} \\
&\alpha_{i,\tau}> 0,\; \forall \; i \in \{1,..,\nue \},\; \label{eqn:alpha_LB} \\
&{\alpha}_{i,\tau}+{\alpha}_{i+1,\tau}=1,\; \forall \; i \in \{1,3,..,\nue-1 \},\; \label{eqn:alpha_sum} \\
&R_{i,\tau} \geq R_{min},\; \forall \; i \in \{1,..,\nue \},\; \label{eqn:cstr}\\ 
&L/2 \geq x_{\uav,\tau} \geq - L/2,\; \\
&L/2 \geq y_{\uav,\tau} \geq - L/2,\;\\
& h_{\uav,\tau} \geq h_0.\;   
\end{align}
\end{subequations}
\begin{remark}
 The problem is feasible if, in addition to the previously stated constraints,  for all $i \in \{1,3,..,\nue-1 \}$, $\alpha_{i,\tau}$
  {satisfies}  the following:
  \vspace{-0.5em}
\begin{equation} \label{eqn:feasible_alpha}
   \alpha_{i,\tau}  \geq  \frac{2^{{R_{min}}/{W}}-1}{SNR_{i,\tau}}, 
\end{equation}
where the received signal to noise ratio at $\tau$ for ${\ue}_i$, $SNR_{i,\tau}$, is expressed as ${P_T \times {g}_{i,\tau}(d_{i,\tau}) \times G} / {\sigma^2}$.~\eqref{eqn:feasible_alpha} follows from~\eqref{eqn: SINR},~\eqref{equ:ith-user_rate}, and~\eqref{eqn:cstr}.
\end{remark}
Solving the non-convex objective function presented in~\eqref{eq:alloc}  is challenging. 
By imposing the condition in~\eqref {eqn:cstr}, we are making sure that the  problem\textquotesingle s solution does not favor the sum rate over fairness, as improving the former may lead to sacrificing servicing some users. 

 {\comm{ Inspired by the success of  GA in solving wireless communications problems and the power of NN in approximating complicated environments, however, we  propose a NEAT-based framework that} To solve this problem, however, we propose AI based framework which trains   the \uav~to 
 \comm{maximize certain objectives,  satisfy requirements, 
adapt to related-unseen environments,
and hence} solve}~\eqref{eq:alloc} efficiently and determine both  the optimal PA  and 3D \uav~placement accordingly.
\vspace{-0.5em}
\section{ P\textnormal{ro}S\textnormal{ky}: A NEAT-based Framework for   UAV 
Network  Resource Management}
\label{sec: ProSky} 
Our proposed framework is designed to have a \uav~environment learned  and modeled using a NN, which as a consequence determines the \uav~3D deployment  and NOMA PA decisions.
In some AI techniques,  DRL, for example,   the NN topology needs to be defined before training, which is usually hard to select and tune for practical applications.   The NN structure needs to be chosen such that the trained model learns the environment well but does not overfit it. During the training process, a NN\textquotesingle s weights and biases are to be optimized to minimize a cost function.  The performance of a NN is affected by the choice of the optimization method. The backpropagation method, for example,  used in DRL,   suffers from the local minima problem. To avoid the aforementioned issues,   we design a novel framework that manages the NOMA-mmWave-\uav~resources based on NEAT.  Unlike many other AI methods, NEAT  evolves and optimizes the parameters of  a number of NNs or solutions simultaneously  using GAs. \comm{ instead of backpropagation.} In addition,    NEAT optimizes and complexifies topologies of NNs to allow complex solutions to evolve. 
In this section,  we provide the NEAT background and  present how we implemented it   to solve the problem on hand effectively. 
\vspace{-1em}
\subsection{NEAT}
In NEAT,\comm{, instead of fixing a NN topology and tuning its parameters,}  a large set of variant-topology NNs, or generations, are tested on a certain task and evaluated based on a certain reward function. Networks that perform well are chosen to contribute, through crossover and mutation, in making a new generation of NNs. Mutation occurs by randomly altering a connection parameters, or adding a node or connection to a NN to ensure diversity of solution and maturity of convergence. The evaluation and evolution processes continue until a certain criterion related to a maximum number of generations, or average reward is met. 
For the effectiveness of NEAT, though, three main concerns are to be taken into account i) disparities and similarities among NNs need to be well represented to crossover  them meaningfully, ii) NNs need not to disappear  prematurely as they could improve performance as they evolve, iii) an effective way that allows NNs to complexify but only as performance demands to avoid slowing down  learning  is required. To handle the {aforementioned} issues, NEAT utilizes some  fundamental techniques. One of which is historical marking. Every NN in NEAT is encoded with two sets. One specifies all nodes, input, output, and hidden,  and another   identifies all connections, their connecting node, weights, enablement status, and an innovation number (IN) assigned uniquely to a connection when created.  In addition, every node is given a  historical marking shared among all NNs  with the same node. Hence, NNs can be tracked, allowing topologies crossover. Another idea of NEAT is speciation. To preserve solutions, NNs are separated  into different species. This allows NNs to compete with their species and solve the issue of eliminating them prematurely.  Complexifying is another  key technique for NEAT.
Topologies in NEAT  develop   incrementally, only as needed, starting from structures with no hidden layers, allowing for finding  {minimal} optimal solutions. We will utilize  these techniques to solve the problem stated in~\eqref{eq:alloc}. \comm{a method that solves the problem stated in~\eqref{eq:alloc}. }
\vspace{-1em}
 \subsection{The P\textnormal{ro}S\textnormal{ky} Model}
 Our proposed framework  is  designed based on NEAT to make the \uav~effectively learn its environment and efficiently take   mobility and  PA actions   such that   fairness and  rate objectives are met. 
The main components of \pro~are described as follows.

\textbf{Initialization.}  To train \pro~effectively, we initialize a  generation of  NNs of size $G_s$  structured from input and output nodes randomly connected.
Input and output  represent system state and \uav~actions respectively, which need to be carefully designed  to determine the most effective NN model.
At any time ${\tau}$ through out the learning process, a state $s_{\tau}$ describes the relative locations of the \uav~to each user,   user\textquotesingle s power coefficient, and \uav-$\ue_i$ channel gain. $s_{\tau}$ is  defined as 
$
s_{\tau}=\big[s_{1,\tau}^T, s_{2,\tau}^T, ..., s_{{\nue},\tau}^T, h_{\uav,\tau}\big]^T$, 
where
$ s_{i,\tau}=\big[\Delta x_{\uav-i,\tau},\Delta y_{\uav-i,\tau},\alpha_{i,\tau},g_{i,\tau}(d_{i,\tau})\big]^T$, $\forall \; i \in \{1,..,\nue \}$. $\Delta x_{\uav-i,\tau}$ and $\Delta y_{\uav-i,\tau}$ are the   x-axis and y-axis difference  between the \uav~and $\ue_i$  locations respectively. The action at ${\tau}$ is defined as
$a_{\tau}=[d_x \delta_x,d_y \delta_y,d_h \delta_h, 
\delta_\alpha^T]^T$ 
 where $\delta_\alpha$ is defined as
$[\delta_{1},\delta_{3}, ...,\delta_{\nue-1}]^T$,  where $\delta_{i}$ is the change in the PA coefficient of  the $\ue_i$. 

\textbf{Evaluation.} The learning process of \pro~occurs over $E$ episodes with $T$ time steps each.  At every time step a system state is fed to a NN, or a \uav, which results in a corresponding action. Based on  the resulted action, the \uav~is given a reward $r_{\tau}$ defined as follows. 
\begin{equation} \label{eq:reward}
\begin{split}
r_\tau = w_r\times \frac{R_{\tau}^{\textrm{tot}}}{W}   \times \prod_{i=1}^{\nue} \mathds{1}_{ \big(\displaystyle R_{i,\tau}\geq R_{min}\big)} + w_s \times\\
\sum_{i=1}^{\nue} \mathds{1}_{\big(\displaystyle R_{i,\tau} \geq R_{min}\big)}
+ w_u \times \sum_{i=1}^{\nue}\frac{R_{i,\tau}}{W} \mathds{1}_{ \big(\displaystyle R_{i,\tau} < R_{min}\big)},
\end{split}
\end{equation}
where   $\mathds{1}_{(.)}$ is the indicator function.   
$w_r$,  $w_s$, and $w_u$, which take non negative  values,  are     total rate,   and satisfied and unsatisfied minimum rate weights  respectively. The  reward is designed, with some similarity to that  proposed in~\cite{adapt}  to facilitate comparison between our NEAT and DRL   frameworks,  carefully  to make the \uav~learn efficiently.
The first  term in~(\ref{eq:reward}) 
aims to increase the total sum rate only if all users meet the minimum rate constraint.
A reward of $w_s$,  with a relatively large value,  contributes to $r_\tau$ to reinforce  satisfying meeting the minimum rate requirement for every user. 
In addition, users which do not satisfy $R_{min}$  result in  unsatisfied minimum rate   reward which is directly proportional to the sum of their rates. In other words, the  \uav~gets rewarded for any improvements it makes for rates that are below $R_{min}$. 
The episode reward for every NN in a generation determines which NNs evolve to the next generation. 
The NNs with the highest reward  are to be included for the next generation. For the rest of  NNs, the higher the average reward, the more the probability it  gets selected for the evolution  process. 

\textbf{Evolution through crossover.}
Historical marking makes a meaningful crossover possible. To crossover two NNs, they get aligned based on their INs. Connections with identical IN, also referred to as matching connections, are randomly selected to appear in the composed NN.  Connections that do not match are called disjoint if they are within the IN range of the other NN  or excess otherwise.  All excess or disjoint connections are included from the NN that achieved a higher average reward  when crossing over.

\textbf{Evolution through mutation.} 
To diversify solutions, mutation, which  either alters existing connections or contribute a new structure to a network, is  implemented. When a new connection is added between two nodes, it is allocated a random 
weight. When a new node   is inserted between two existing connected nodes.  While the link between the previous start  and new nodes keeps the old connection parameters,  the other  link is assigned a weight of $1$.

\textbf{Solution preservation.}
This component enables, through speciation, novel topologies, which may initially perform poorly, to be constructed and optimized without the worry of being destroyed before they can be fully investigated. Speciation, basically, divides a generation into several species based on topological and connectivity similarities. If a compatibility distance, 
{which is  proportional  to the weighted sum of the number of excesses, disjoints, and the variations in weights of matching connection, }
  is less than a compatibility threshold $\delta_{th}$, a two  NNs are said to belong to the same species~\cite{stanley2002evolving}. 
  \comm{$\delta$ is defined in~\cite{stanley2002evolving} as
 \begin{equation}
    \delta = \frac{c_e N_e}{N_c} + \frac{c_d N_d}{N_c} + c_m \cdot \bar{W_{m}},
\end{equation}
where  $N_e$ and $N_d$ are  the number of excess and disjoint links respectively.  $\bar{W_{m}}$ is the average weight differences of matching connections.  $N_c$ is the largest total number of connections among the two NNs.
$c_e, c_d$, and $c_m$ 
are  the coefficients that determine the impact of the dissimilarity between two NNs in terms  of their disjoint, excess, and the variation  in weights of matching connection. }
 A number of species are created in the first episode and sequentially ordered. At subsequent episodes, a random NN is selected to represent each species. Any other generated NN  is assigned to the first species that  is  compatible with its representative NN. A new species is created in case there is no compatibility.  

\textbf{Solutions reduction.} 
To avoid having a species  takes over the entire generation by getting too big, an adjusted reward function is introduced~\cite{stanley2002evolving}. The idea is that the reward of every solution in a species gets normalized by the number of NNs belongs to the species.  Each species gets assigned  only number of NNs proportional to the sum of the  adjusted reward of corresponding species members. That as a consequence may result in eliminating the solutions with the lowest performance and reduce number of solutions within a species. 

\comm{ \blue{
 \textbf{\pro~relation with NEAT.} 
 In \pro, the NN will represent a \uav~that takes decisions according to each state given at certain time. At the beginning of the first episode, a total of $G_s$ NNs are created with minimal structure (without any hidden nodes) and with random weights and biases. Each NN will take the same initial state at the first time step $\tau = 0$, however, each NN will choose a certain actions that will lead them to different states which depends on the initial NN randomization occurred at the beginning of the episode. At each time step, we will give each NN a reward according to equation \ref{eq:reward}, at the end of the episode (i.e. after $T$ time steps), we will average the total rewards given for each NN and we will use it to evaluate them. The NN with the best reward will be chosen to contribute to the next episode directly, and as discussed above, the better the NN reward, the highest the probability it will get chosen for the evolution process. Afterwards, the selected NNs will be evolved through crossover and mutations, and then speciation process will occur. Finally these new NNs together with the best NNs chosen from the last episode will contribute to the next episode which will lead to different solutions.
 }}
 
 {The last four components of \pro, which  are mainly based on NEAT~\cite{stanley2002evolving}, determine how NNs in \pro~are evolved. However, the initialization and evaluation components are    designed by us  to get the most out of NEAT and make it   learn the 3D network environment effectively  and manage    resources  efficiently as we will show in the next section.} 
\vspace{-0.5em}
\section{P\textnormal{ro}S\textnormal{ky} Performance Evaluation}
\label{sec: numr}

 \subsection{Implementation Settings}
\pro~is trained for $1000$ episodes, or generations, with $300$ time steps each  {and $G_s$ of $50$} NNs which are assumed to  be  feed-forward,  fully-connected at the initialization, and activated using ReLU.
Weight and biases are generated from standard normal in a range of $[-30,30]$ with a {mutation rate $0.8$ and $0.7$ respectively}. A node, and connection, add or delete probabilities are all  set to $0.2$. 
 $\delta_{th}$ is chosen to be $3$. 
 
 The  proposed framework is simulated  for a $100 \times 100$ $m^2$ urban area  with $\nue =4$. 
Users are located, relative to the center of $\mathcal{A}$,  at $(4,15)$, $(-44,-49)$, $(-5,21)$ and $(47,49)$. 
The  \uav~is assumed to keep a minimum height of $ 10$ m,  and, at the beginning of each {episode}, to be deployed at       $(0,0,50)$  and  assign $0.5 P_T$ for each user. The  change  in  the  PA coefficient is set to $\pm 0.01$  for all users. $\delta_x$, $\delta_y$, and $\delta_{h}$ are all   set to be $1$ m. 
%
%
Channel is modeled according to~\cite{akdeniz2014millimeter}, 
 where the path loss intercept  and  exponent  are set to $10^{-6.4}$ and $2$ respectively. 
 Thermal noise power $\sigma^2$ is assumed to be $-84$ dBm. \comm{ Unless otherwise stated,..EE} The transmit power $P_T$ is set to $20$ dBm. 
 Antenna configurations $\mathcal{N}_{UAV} \times \mathcal{N}_{UE}$ are chosen to be $8 \times  8$. System bandwidth $W$ and carrier frequency $f_c$  are taken $2$ GHz and $28$ GHz respectively. 
 \vspace{-2em}
\subsection{Performance Analysis}

\label{subsec:Numerc_analysis}
In this subsection, we provide the performance  of \pro~in managing 3D NOMA-\uav~network  both in training and  testing scenarios.
We compare our finding with the state-of-art technique provided in~\cite{Chen2019}, 
which throughout the section will be referred to as SoA. In~\cite{Chen2019}, the authors solved the non-convex NOMA PA and \uav~placement problem, disjointly while  restricting the \uav~placement to a 2D plane, set at a height of $50$ in our analysis,  using the conventional optimization framework. 

  In addition, we compare the \pro~performance with the DRL based algorithm \adapt, not just because we are  examining the performance of the two different AI tools these algorithms are  built on, but also because \adapt, for the best of our knowledge, is the only AI based framework proposed to solve the 3D \uav~placement and NOMA PA jointly. 

For  evaluation, we introduce the performance metrics  ${R}_{e}^{\textrm{tot}}$ defined as achieved average sum-rate per generation  in case of \pro~and by averaging the average rate  per episode and over the most recent $100$ episodes for \adapt. 
${R}_{e}^{\textrm{tot}}$ \adapt~equals zero for all episodes $ep < 100$.  The average total rate varies drastically from episode to another in case of \adapt, hence averaging over  $100$ episodes was needed to smooth out the data. Using  some performance metrics, including ${R}_{e}^{\textrm{tot}}$, we evaluate \pro~training and testing performance in the following aspects:

\begin{figure}[t]
    \centering
    \includegraphics[width=0.385\textwidth]{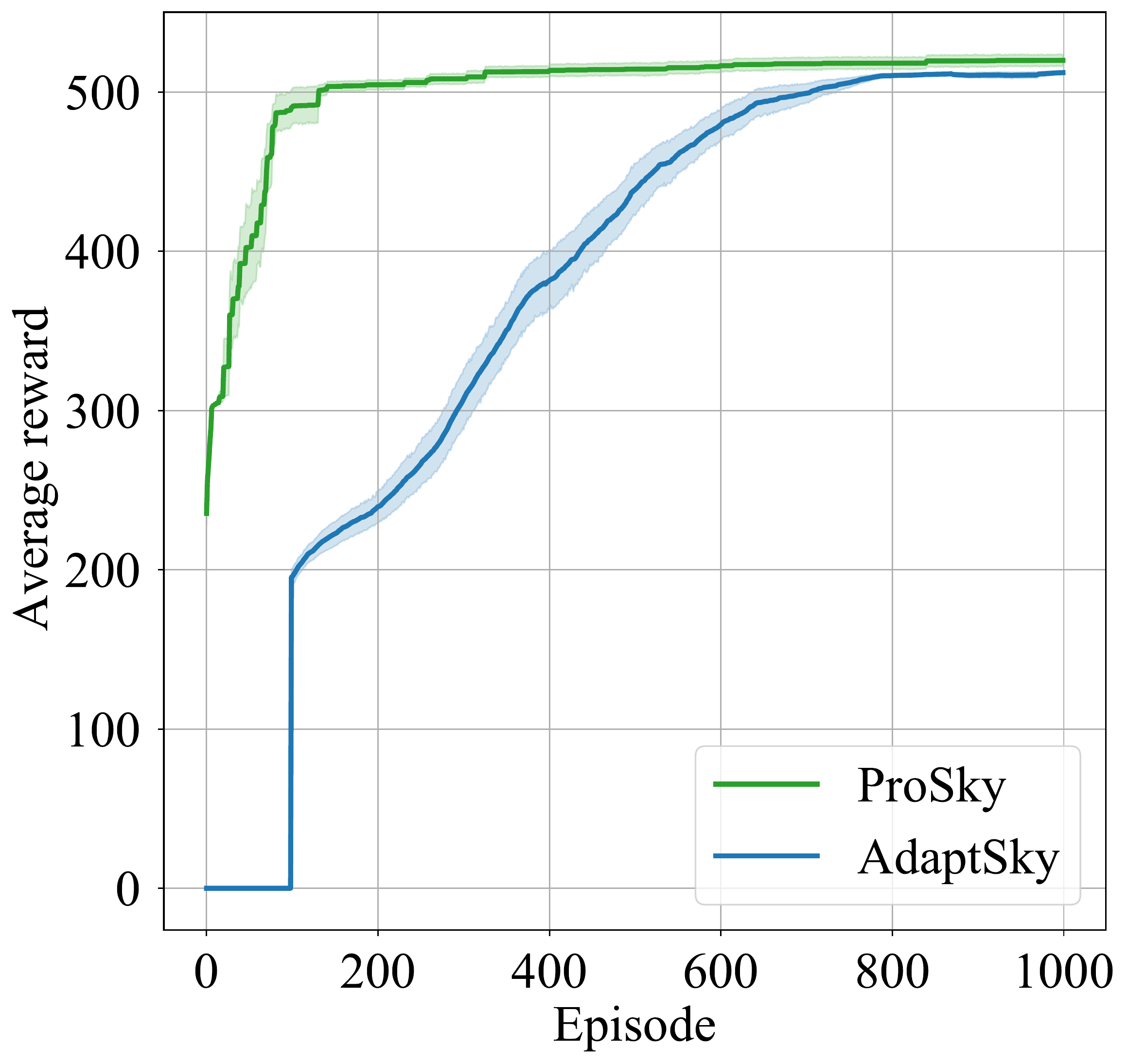}
    \caption{Episode training average rewards.} 
    \label{fig: ProSky-VS-AdaptSky-SUMRATE-CI}
\end{figure}

  \textbf{Learning rate.}
  %
%
In order to show how efficient and fast \pro~can handle the problem in hand, we, in 
Fig. \ref{fig: ProSky-VS-AdaptSky-SUMRATE-CI}, show the average reward for  \pro~and AdaptSky,  as they both learn the \uav~environment. 
The average rewards is defined similar to  ${R}_{e}^{\textrm{tot}}$, i.e., it is the episode and $100$ episodes reward average for \pro and \adapt~respectively. 
The training parameters are set as ${R}_{min}/W = 0.5$, $w_s = 100$, $w_r = 1$, 
and $w_u = 1$.
To demonstrate the consistency of our proposed method, the figure includes the   confidence interval (CI) of one standard deviation of $10$ different simulation runs. 
Interestingly,  \pro~achieves $96.17$\% of the average reward convergence value more than $5.3$ times faster than that for \adapt, 
even though \adapt~is built based on  dueling architecture, which  advances   DRL learning rate{~\cite{wang2016dueling}}. This speedup rate is   $96\%$ higher than  the average, and only $2\%$ off of the highest,  achieved end-to-end speedup rate  of ActorQ network~\cite{krishnanquarl} which leverages low-precision quantized actors to speed up the  learning process.

Furthermore, based on Fig.~\ref{fig: ProSky-VS-AdaptSky-SUMRATE-CI},  \pro~converges to a slightly better reward, or total sum data rate, than \adapt~with  less uncertainty. 
\comm{
%
}

    \begin{figure}[t]
    \centering
    \includegraphics[width=0.385\textwidth]{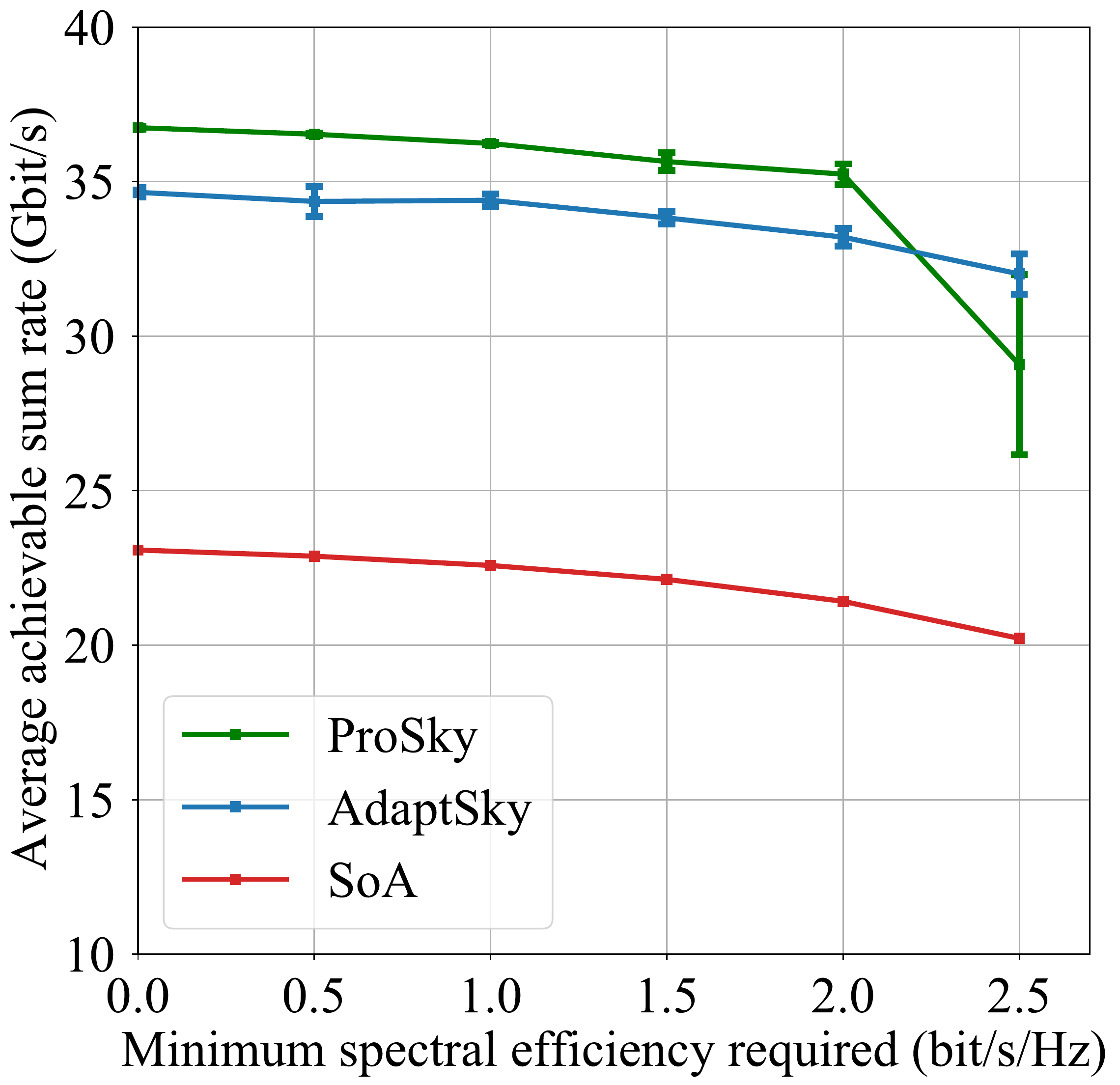}
    \caption{Average sum rate for different fairness requirements.}
    \label{fig: min-sum-vs-se}
\end{figure}
    \textbf{Rate and fairness (training).}
     %
    %
    In this scenario, we vary  the minimum  SE required  ${R}_{min}/W$, as a way to impose fairness among users, 
    and observe   
    ${R}_{e}^{\textrm{tot}}$. We   depict  the convergence value of ${R}_{e}^{\textrm{tot}}$ over a thousand training episodes
    in Fig. \ref{fig: min-sum-vs-se} and show the CI  of $5$ runs.   
    The reward function in this scenario is structured  to give a higher reward whenever a user exceeds ${R}_{min}$, which can be achieved through setting the rewards parameters as   $w_r= 10$,   {$w_s= 100$}, and $w_u = 1$.  
    We can observe from Fig. \ref{fig: min-sum-vs-se}  that \pro~outperforms SoA,  very consistently i.e. with tight CI, and  improvement of up to $59.23\%$.  
In addition,  when the minimum SE is less than or equal $2$ bit/s/Hz, \pro~outperforms \adapt~with average  of  {$5.85$\%} which corresponds to $1.99$ Gbit/s improvement.  
The uncertainty at a higher SE minimum requirement is expected as \pro~finds the best model by parallelly trying number of random solutions which may lead to a lower average performance than \adapt, as  feasible solutions are getting very limited.

\begin{figure}[t]
    \centering
    \includegraphics[width=0.38\textwidth]{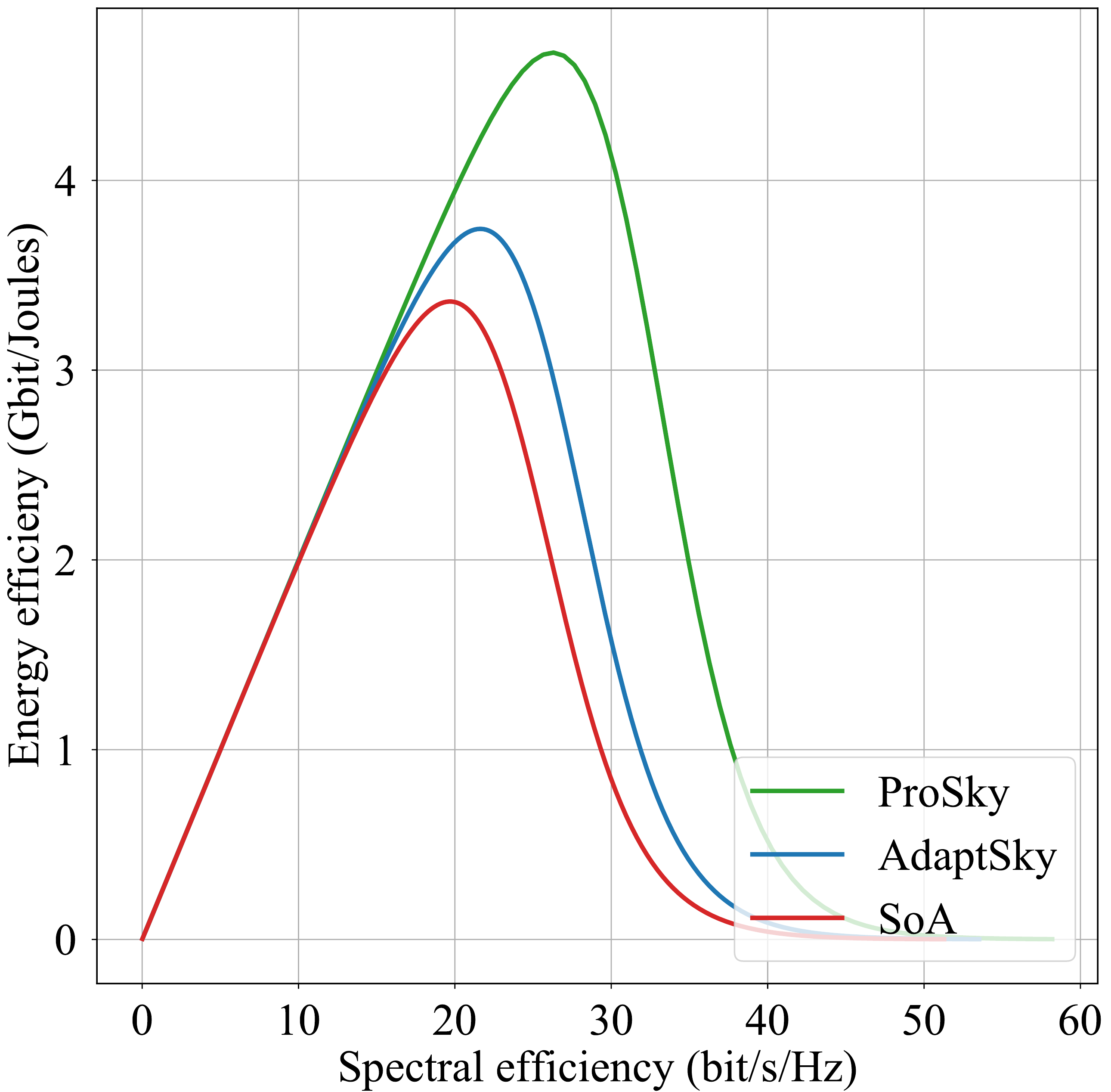}
    \caption{ Energy efficiency performance during testing.}
    \label{fig: ProSky-vs-SoA-EE}
\end{figure}

\textbf{Energy efficiency (testing).}
In this scenario, we train \pro~to optimize the sum rate objective with a setup similar to that in the learning rate  scenario, but with $R_{min}$=0.
The best NN model  from  training is imported and tested for every power in the range  $-20$ dBm to $80$ dBm with a step size of $0.1$ dBm,  over $300$ time steps through which the \uav~interacts with the observed states. We determine the average SE over all time steps and calculate the corresponding EE, defined as SE divided by power required for signal transmission plus some other consumed power assumed to take the value of $40$ dBm.    
  In Fig. \ref{fig: ProSky-vs-SoA-EE}, we plot the EE for \pro~and the two baselines, which  \pro~obviously outperforms. At the green point, the highest EE point in the curves, \pro~shows an improvement of $21.93$\% and $35.71$\% over \adapt~and SoA respectively.

  \pro~exhibits a fast adapting learning model and offers, while meeting  fairness requirements, huge gain in terms of SE and EE over the {model-based} solution proposed in~\cite{Chen2019}, which seem to fall short in handling  complicated networks. Similarly, \pro~demonstrates some improvement over the  DRL-based algorithm proposed in~\cite{adapt} in terms of  SE and EE, and significant enchantment in learning rate.  
 
\vspace{-1em}
\section{Conclusion and Discussion}
\label{sec: conc}
In this paper, we  proposed \pro, a novel AI-based framework built on NEAT algorithm which merges GA and deep learning. 
 \pro~optimizes the \uav~3D location while allocating NOMA resources in the   mmWave spectrum such that  fairness is guaranteed and the users\textquotesingle~sum rate is maximized.
 \pro~significantly outperforms a model-based scheme, in terms of SE and EE. 
 Moreover, \pro~has shown a huge gain in learning speed and some gains in SE and EE over a DRL-based baseline. 
  To get the most out of integrating GA and deep learning, nonetheless, and to  draw a more comprehensive performance analysis,  more studies about the applicability of NEAT and its advances in more generic channel models in   multi-UAV networks are encouraged considering  the improvement our findings  {exhibited} in resources management. We believe NEAT will require more investigations in terms of the NN evolution and  selection process in such cases.
   
\vspace{-0.5em}
\bibliographystyle{IEEEtran}
\bibliography{ProSky}

\end{document}